\documentclass{ws-procs10x7}

\def\lsim{\raise0.3ex\hbox{$<$\kern-0.75em\raise-1.1ex\hbox{$\sim$}}}
\def\gsim{\raise0.3ex\hbox{$>$\kern-0.75em\raise-1.1ex\hbox{$\sim$}}}
\def\noi{\noindent}
\def\nn{\nonumber}
\def\bea{\begin{eqnarray}}  \def\eea{\end{eqnarray}}
\def\beq{\begin{equation}}   \def\eeq{\end{equation}}
\begin{document}

\title{Sum rules in the heavy quark limit of QCD \\ and Isgur-Wise functions}

\author{F. Jugeau, A. Le Yaouanc, L. Oliver and J.-C. Raynal}

\address{Laboratoire de Physique Th\'eorique, Unit\'e Mixte de Recherche
UMR 8627 - CNRS, Universit\'e de Paris XI, B\^atiment 210, 91405
Orsay Cedex, France\\E-mail: luis.oliver@th.u-psud.fr}

\address{\it Presented by L. Oliver}

\twocolumn[\maketitle\abstract{Using the OPE, we formulate new sum 
rules in the heavy quark limit of
QCD. These sum rules imply that the elastic Isgur-Wise function $\xi
(w)$ is an alternate series in powers of $(w-1)$. Moreover, one gets
that the $n$-th derivative of $\xi (w)$ at $ w=1$ can be bounded by the
$(n-1)$-th one, and an absolute lower bound for the $n$-th derivative
$(-1)^n \xi^{(n)}(1) \geq {(2n+1)!! \over 2^{2n}}$.  Moreover, for the
curvature we find $\xi ''(1) \geq {1 \over 5} [4 \rho^2 + 3(\rho^2)^2]$
where $\rho^2 = - \xi '(1)$. We  show that the quadratic term ${3 \over
5} (\rho^2)^2$ has a  transparent physical interpretation, as it is
leading in a non-relativistic expansion in the mass of the  light
quark. These bounds should be taken into account in the
parametrizations of $\xi (w)$ used to extract $|V_{cb}|$. These results
are consistent with the dispersive bounds, and they strongly reduce the
allowed region of the latter for $\xi (w)$. The method is extended to
the subleading quantities in $1/m_Q$, namely $\xi_3(w)$ and
$\overline{\Lambda}\xi (w)$.}]

In the leading order of the heavy quark expansion of QCD, Bjorken sum
rule (SR) \cite{1r} relates the slope of the elastic Isgur-Wise (IW)
function $\xi (w)$, to the IW functions of the transitions between the
ground state and the $j^P = {1 \over 2}^+$, ${3 \over 2}^+$ excited
states, $\tau_{1/2}^{(n)}(w)$, $\tau_{3/2}^{(n)}(w)$, at zero recoil $w
= 1$ ($n$ is a radial quantum number). This SR leads to the lower bound
$- \xi '(1) = \rho^2 \geq {1 \over 4}$. Recently, a new SR was
formulated by Uraltsev in the heavy quark limit \cite{2r} involving
also $\tau_{1/2}^{(n)}(1)$, $\tau_{3/2}^{(n)}(1)$, that implies,
combined with Bjorken SR, the much stronger lower bound $\rho^2 \geq {3
\over 4}$, a result that came as a big surprise. In ref. \cite{3r}, in
order to make a systematic study in the heavy quark limit of
QCD, we have developed a manifestly covariant formalism within the
Operator Product Expansion (OPE). We did recover Uraltsev SR plus a new
class of SR. Making a natural physical assumption, this new class of SR
imply the bound $\sigma^2 \geq {5 \over 4} \rho^2$ where $\sigma^2$ is
the curvature of the IW function. Using this formalism including the
whole tower of excited states $j^P$, we have recovered rigorously the
bound $\sigma^2 \geq {5 \over 4} \rho^2$ plus generalizations that
extend it to all the derivatives of the IW function $\xi (w)$ at zero
recoil, that is shown to be an alternate
series in powers of $(w-1)$. \par

Using the OPE and the trace formalism in the heavy quark limit,
different initial and final four-velocities $v_i$ and $v_f$, and heavy
quark currents, where $\Gamma_1$ and $\Gamma_2$ are arbitrary Dirac
matrices $J_1 = \bar{h}_{v'}^{(c)}\ \Gamma_1\
h_{v_i}^{(b)}$, $J_2 = \bar{h}_{v_f}^{(b)}\ \Gamma_2\
h_{v'}^{(c)}$, the following sum rule can be written
\cite{4r}~: \bea \label{2e} &&\Big \{ \sum_{D=P,V} \sum_n Tr \left [
\bar{\cal B}_f(v_f) \bar{\Gamma}_2 {\cal D}^{(n)}(v') \right ] \nn \\
&&Tr \left [ \bar{\cal D}^{(n)}(v') \Gamma_1 {\cal B}_i(v_i)\right ]
\xi^{(n)} (w_i) \xi^{(n)} (w_f) \nn \\ &&+ \ \hbox{Other excited
states} \Big \}  = - 2 \xi(w_{if}) \nn \\ &&Tr \left [ \bar{\cal
B}_f(v_f) \bar{\Gamma}_2 P'_+ \Gamma_1 {\cal B}_i(v_i)\right ]\ .\eea

In this formula $v'$ is the intermediate meson four-velocity, $P'_+ =
{1 \over 2} (1 + {/ \hskip - 2 truemm v}')$ comes from the residue of
the positive energy part of the $c$-quark propagator, $\xi(w_{if})$ is
the elastic Isgur-Wise function that appears because one assumes $v_i
\not= v_f$. ${\cal B}_i$ and ${\cal B}_f$ are the $4 \times 4$ matrices
of the ground state $B$ or $B^*$ mesons and ${\cal D}^{(n)}$ those of
all possible ground state or excited state $D$ mesons coupled to $B_i$
and $B_f$ through the currents. In (\ref{2e}) we
have made explicit the $j = {1 \over 2}^-$ $D$ and $D^*$
mesons and their radial excitations of quantum number $n$. The explicit
contribution of the other excited states is written below.\par

The variables $w_i$, $w_f$ and $w_{if}$ are defined as $w_i = v_i 
\cdot v'$, $w_f = v_f \cdot v'$, $w_{if} = v_i \cdot
v_f$. \par

The domain of $(w_i$, $w_f$, $w_{if}$) is \cite{3r} ($w_i, w_f \geq 1$)
\bea \label{4e} &&w_iw_f - \sqrt{(w_i^2 - 1)
(w_f^2 - 1)} \leq w_{if} \nn \\
&&\leq w_iw_f + \sqrt{(w_i^2 -1) ( w_f^2 - 1)} \
. \eea
The SR (\ref{2e}) writes $L\left ( w_i, w_f, w_{if} \right ) = R \left ( w_i,
w_f, w_{if} \right )$, where $L(w_i, w_f, w_{if})$ is the sum over 
the intermediate
charmed states and $R(w_i, w_f, w_{if})$ is the OPE side. Within the
domain (\ref{4e}) one can derive relatively to any of the variables
$w_i$, $w_f$ and $w_{if}$ and obtain different SR taking different 
limits to the frontiers
of the domain. \par

As in ref. \cite{3r}, we choose as initial and final states the $B$
meson ${\cal B}_i (v_i) = P_{i+} (- \gamma_5)$ ${\cal B}_f (v_f) = 
P_{f+} (- \gamma_5)$ and vector or axial currents projected along the 
$v_i$ and $v_f$
four-velocities
\beq \label{8e} J_1 = \bar{h}_{v'}^{(c)}\ {/ \hskip - 2 truemm v}_i\
h_{v_i}^{(b)} \quad , \qquad J_2 = \bar{h}_{v_f}^{(b)}\ {/ \hskip - 2
truemm v}_f\ h_{v'}^{(c)}  \eeq
\noi we
obtain SR (\ref{2e}) with the sum of all excited states $j^P$ in a
compact form~:
$$(w_i + 1) (w_f + 1) \sum_{\ell \geq
0} {\ell + 1 \over 2 \ell + 1} S_{\ell} (w_i, w_f, w_{if})$$
$$\sum_n \tau_{\ell + 1/2}^{(\ell)(n)}(w_i) \tau_{\ell + 1/2}^{(\ell
)(n)}(w_f)$$ $$+ \sum_{\ell \geq 1} S_{\ell} (w_i, w_f, w_{if})
\sum_n \tau_{\ell - 1/2}^{(\ell)(n)}(w_i) \tau_{\ell - 1/2}^{(\ell
)(n)}(w_f)$$ \beq \label{9e}= (1 + w_i+w_f+w_{if}) \xi(w_{if}) \ .\eeq

We get, choosing instead the axial currents,
\beq \label{10e} J_1 = \bar{h}_{v'}^{(c)}\ {/ \hskip - 2 truemm v}_i\
\gamma_5 \ h_{v_i}^{(b)} \ , \ J_2 = \bar{h}_{v_f}^{(b)}\ {/
\hskip - 2 truemm v}_f\ \gamma_5\ h_{v'}^{(c)} \ , \eeq
\bea
\label{11e}
  &&\sum_{\ell \geq 0} S_{\ell + 1}
(w_i, w_f, w_{if}) \nn \\
&&\sum_n \tau_{\ell + 1/2}^{(\ell)(n)}(w_i) \tau_{\ell
+ 1/2}^{(\ell )(n)}(w_f)\nn \\
&&+ (w_i - 1) (w_f - 1) \nn \\
&&\sum_{\ell \geq
1} {\ell \over 2 \ell - 1} S_{\ell - 1} (w_i, w_f, w_{if})\nn \\
&&\sum_n
\tau_{\ell - 1/2}^{(\ell)(n)}(w_i) \tau_{\ell - 1/2}^{(\ell )(n)}(w_f)\nn \\
&&= - (1 - w_i-w_f+w_{if}) \xi (w_{if}) \ .\eea

Following the formulation of heavy-light states for arbitrary $j^P$
given by Falk \cite{4r}, we have defined in ref. \cite{3r} the IW
functions $\tau_{\ell + 1/2}^{(\ell)(n)}(w)$ and $\tau_{\ell -
1/2}^{(\ell)(n)}(w)$, $\ell$ and $j = \ell \pm {1
\over 2}$ being the orbital and total angular momentum
of the light quark. \par

In (\ref{8e}) and (\ref{10e}) $S_n$ is given by
\bea \label{12e} &&S_n = v_{i\nu_1} \cdots v_{i\nu_n}\ v_{f\mu_1} \cdots
v_{f\mu_n} \nn \\
&&\sum_{\lambda} \varepsilon'^{(\lambda )*\nu_1 \cdots \nu_n} \
\varepsilon'^{(\lambda )\mu_1 \cdots \mu_n} \ .\eea
\noi One can show \cite{3r}~:
\bea \label{14e} &&S_n = \sum_{0 \leq k \leq {n \over 2}}
C_{n,k} (w_i^2 - 1)^k (w_f^2 - 1)^k \nn \\
&&(w_i w_f - w_{if})^{n-2k} \eea
\noi with $C_{n,k} = (-1)^k {(n!)^2 \over (2n) !} \ {(2n - 2k) !
\over k! (n-k) ! (n-2k)!}$.\par

\noi From the sum of (\ref{9e}) and (\ref{11e}) one obtains, 
differentiating relatively to
$w_{if}$ \cite{5newr} $(\ell \geq 0)$~:

$$\xi^{(\ell
)} (1) = {1 \over 4} \ (-1)^{\ell} \ \ell ! \left \{ {\ell + 1 \over 2
\ell + 1} 4 \sum_n \left [ \tau_{\ell + 1/2}^{(\ell )(n)}(1) \right
]^2 \right .$$
\beq \label{16e}
\left . + \sum_n \left [ \tau_{\ell - 1/2}^{(\ell -1)(n)}(1) \right
]^2 + \sum_n \left [ \tau_{\ell - 1/2}^{(\ell)(n)}(1) \right ]^2 \right
\}   \ .\eeq
\noi This relation shows that $\xi (w)$ is an alternate series in
powers of $(w-1)$. Equation (\ref{16e}) reduces to Bjorken SR \cite{1r}
for $\ell = 1$. Differentiating (\ref{11e}) relatively to $w_{if}$ and
making $w_i=w_f = w_{if} = 1$ one obtains~:
\beq \label{17e} \xi^{(\ell )} (1) = \ell !\
(-1)^{\ell} \sum_n \left [ \tau_{\ell + 1/2}^{(\ell)(n)}(1) \right ]^2
\quad (\ell \geq 0) \ . \eeq
\noi Combining (\ref{16e}) and (\ref{17e}) one obtains a SR for all 
$\ell$ that reduces to Uraltsev SR
\cite{2r} for $\ell = 1$. From (\ref{16e}) and (\ref{17e}) one obtains~:
  $$(-1)^{\ell} \ \xi^{(\ell)} (1) = {1 \over 4} \ {2 \ell
+ 1 \over \ell} \ell !$$
\beq \label{19e}
\left \{ \sum_n \left [ \tau_{\ell - 1/2}^{(\ell
- 1)(n)}(1) \right ]^2 + \sum_n \left [ \tau_{\ell -
1/2}^{(\ell)(n)}(1) \right ]^2 \right \} \ . \eeq
\noi implying
$$(-1)^{\ell} \xi^{(\ell)}(1) \geq  {2\ell + 1 \over 4}
\left [  (-1)^{\ell - 1} \xi^{(\ell - 1)}(1) \right ] $$
\beq\label{20e} \geq  {(2\ell + 1)!!
\over 2^{2\ell}} \eeq
\noi that gives, in particular, for the lower cases,
\beq
\label{22e}
- \xi ' (1) = \rho^2 \geq {3 \over 4} \quad , \quad \xi '' (1) \geq 
{15 \over 16}
\eeq

Considering systematically the derivatives of the SR (\ref{9e}) and 
(\ref{11e}) relatively to $w_i$, $w_f$, $w_{if}$ with the boundary 
conditions $w_{if}
= w_i = w_f = 1$, one obtains a new SR:
\beq
\label{25e}
{4 \over 3} \rho^2 + (\rho^2)^2 - {5 \over 3} \sigma^2 + \sum_{n\not=
0} |\xi^{(n)'}(1)|^2 = 0
\eeq
\noi that implies~:
   \beq
\label{27e}
\sigma^2 \geq {1 \over 5} \left [ 4 \rho^2 + 3(\rho^2)^2 \right ] \ .
\eeq
There is a simple intuitive argument to understand the term ${3 \over
5} (\rho^2)^2$ in the best bound (\ref{27e}), namely the
non-relativistic quark
model, i.e. a non-relativistic light quark $q$ interacting with a
heavy quark $Q$ through a potential. The form factor has the simple form~:
\bea
\label{28e}
&&F({\bf k}^2)= \int d {\bf r} \ \varphi^+_0(r)\nn \\
&& \exp \left ( i {m_q
\over m_q + m_Q} {\bf k} \cdot {\bf r} \right ) \varphi_0 (r)
\eea
\noi where $\varphi_0(r)$ is the ground state radial wave function. 
Identifying the non-relativistic IW function $\xi_{NR}(w)$ with
the form factor $F({\bf k}^2)$ (\ref{28e}),  one can prove that,
\beq
\label{31e}
\sigma_{NR}^2 \geq {3 \over 5} \ \left  [ \rho^2_{NR} \right ]^2 \ .
\eeq

Thus, the non-relativistic limit is a good guide-line to study the
shape of the IW function $\xi (w)$. We have recently generalized the
bound (\ref{31e}) to all the derivatives of $\xi_{NR}(w)$. The method uses
the positivity of matrices of moments of the ground state wave function
\cite{6newr}. We have shown that the method can be generalized to the real
function $\xi(w)$ of QCD.\par

An interesting phenomenological remark is that the simple
parametrization for the IW function \cite{5r}
\beq
\label{32e}
\xi (w) = \left ( {2 \over w + 1} \right )^{2 \rho^2}
\eeq
\noi satisfies the inequalities (\ref{20e}), (\ref{27e}) if
$\rho^2 \geq {3 \over 4}$.
\par

The result (\ref{20e}), that shows that all derivatives at zero recoil
are large, should have important phenomenological implications for the
empirical fit needed for the extraction of $|V_{cb}|$ in $B \to D^*\ell
\nu$.  The usual fits to extract $|V_{cb}|$
using a linear or linear plus quadratic dependence of $\xi (w)$ are not
accurate enough. It is
important to point out that the most precise data points are the ones
at large $w$, so that higher derivatives contribute importantly in
this region. \par

A considerable effort has been developed to formulate dispersive
constraints on the shape of the form factors in $\bar{B} \to D^*\ell
\nu$ \cite{7r}-\cite{8r}, at finite mass. \par

Our approach, based on Bjorken-like
SR, holds {\it in the physical region} of the semileptonic decays
$\bar{B} \to D^{(*)}\ell \nu$ and
{\it in the heavy quark limit}. The two approaches are quite 
different in spirit and in their
results. \par

Let us consider the main results of ref. \cite{8r}
summarized by the one-parameter formula
\beq
\label{35e}
\xi(w)  \cong 1 - 8 \rho^2 z + (51\rho^2 - 10)z^2 -
(252 \rho^2 - 84) z^3
\eeq
\noi with the variable $z(w)$ defined by
\beq
\label{36e}
z = {\sqrt{w+1} - \sqrt{2}\over \sqrt{w+1} + \sqrt{2}}
\eeq
\noi and the allowed range for $\rho^2$ being $- 0.17 < \rho^2 < 
1.51$. This domain is considerably tightened by the lower bound on
$\rho^2$~: ${3 \over 4} \leq \rho^2 < 1.51$, that shows that our type 
of bounds are complementary to the
upper bounds obtained from dispersive methods.\par

By extension of our method to subleading order in $1/m_Q$, we have
shown that the subleading quantities, that are functions of $w$,
$\overline{\Lambda}\xi (w)$ and $\xi_3(w)$ can be expressed in terms of leading
quantities, namely the ${1\over 2}^- \to {1 \over 2}^+, {3 \over 2}^+$
IW functions $\tau_j^{(n)}(w)$ and the corresponding level spacings
$\Delta E_j^{(n)}$ $(j = {1 \over 2}, {3 \over 2})$ \cite{9r}
\bea \label{44e} \overline{\Lambda} \xi (w) &=& 2(w+1) \sum_n \Delta
E_{3/2}^{(n)} \ \tau_{3/2}^{(n)}(1) \ \tau_{3/2}^{(n)}(w)\nn \\ &&+ \ 2
\sum_n \Delta E_{1/2}^{(n)} \ \tau_{1/2}^{(n)}(1) \ \tau_{1/2}^{(n)}(w)
\eea
\bea \label{45e} \xi_3 (w) &=& (w+1) \sum_n \Delta E_{3/2}^{(n)} \
\tau_{3/2}^{(n)}(1) \ \tau_{3/2}^{(n)}(w)\nn \\ &&- \ 2 \sum_n \Delta
E_{1/2}^{(n)} \ \tau_{1/2}^{(n)}(1) \ \tau_{1/2}^{(n)}(w) \ . \eea

\noi These quantities reduce to known SR for $w=1$, respectively
Voloshin SR \cite{10r} and a SR for $\xi_3(1)$ \cite{11r,2r}, and
generalizes them for all $w$.\par

The comparison of (\ref{44e}), (\ref{45e}) with the results of the BT 
quark model
\cite{5r} is very encouraging. Within this scheme $\xi (w)$ is given by
(\ref{32e}) with $\rho^2 = 1.02$, while one gets, for the $n = 0$ states
\beq \label{65} \tau_{j}^{(0)}(w) = \tau_{j}^{(0)}(1)\left ( {2
\over w+1}\right )^{2\sigma_{j}^2}
\eeq

\noi with $\tau_{3/2}^{(0)}(1) = 0.54$, $\sigma_{3/2}^2 = 1.50$,
$\tau_{1/2}^{(0)}(1) = 0.22$ and $\sigma_{1/2}^2 = 0.83$. Assuming the
reasonable saturation of the SR with the lowest $n = 0$ states
\cite{5r}, one gets, from the first relation (\ref{44e}), a sensibly
constant value for $\overline{\Lambda} = 0.513 \pm 0.015$. \par

In conclusion, using sum
rules in the heavy quark limit of QCD, as formulated in ref.
\cite{3r,9r}, we
have found lower bounds for the moduli of the derivatives of $\xi (w)$. Any
phenomenological parame\-trization of $\xi (w)$ intending to fit the
CKM matrix element $|V_{cb}|$ in $B \to
D^{(*)}\ell \nu$ should satisfy
these bounds. Moreover, we discuss these bounds in comparison with 
the dispersive
approach. We show that there is no contradiction, our bounds
restraining the region for $\xi (w)$ allowed by this latter method. 
Moreover, we have found non-trivial new information on subleading 
contributions in $1/m_Q$.

\section*{Acknowledgement.} We are indebted to the EC contract
HPRN-CT-2002-00311 (EURIDICE) for its support.


\begin{thebibliography}{99}

\bibitem{1r} J. D. Bjorken, invited talk at Les Rencontres de la
Vall\'ee d'Aoste, La Thuile, SLAC-PUB-5278, 1990~; N. Isgur and M. B.
Wise, Phys. Rev. {\bf D43}, 819 (1991).


\bibitem{2r} N. Uraltsev, Phys. Lett. {\bf B501}, 86 (2001).

\bibitem{3r} A. Le Yaouanc, L. Oliver and J.-C. Raynal, Phys. Rev. 
{\bf D67}, 114009 (2003).

\bibitem{4r} A. Falk, Nucl. Phys. {\bf B378}, 79 (1992).

\bibitem{5newr} A. Le Yaouanc, L. Oliver and J.-C. Raynal, Phys. 
Lett. {\bf B557}, 207 (2003).

\bibitem{6newr} F. Jugeau, A. Le Yaouanc, L. Oliver and J.-C. Raynal, hep-ph/0405234, to appear in Phys. Rev. D.

\bibitem{5r} V. Mor\'enas, A. 
Le Yaouanc, L. Oliver, O. P\`ene and
J.-C. Raynal, Phys. Rev. {\bf D56}, 5668 (1997).

\bibitem{7r} C. G. Boyd, B. Grinstein and R. F. Lebed, Phys. Rev. {\bf
D56}, 6895 (1997).

\bibitem{8r} I. Caprini, L. Lellouch and M. Neubert, Nucl. Phys. {\bf
B530}, 153 (1998).

\bibitem{9r} J. Jugeau, A. Le Yaouanc, L. Oliver and J.-C. Raynal, 
hep-ph/0407176, to appear in Phys. Rev. D.

\bibitem{10r} M. Voloshin, Phys. Rev. {\bf D46}, 3062 (1992).

\bibitem{11r} A. Le Yaouanc et al., Phys. Lett. {\bf B480}, 119 (2000).

 


  \end{thebibliography}
\end{document}